OPEN

# Superconductivity with extremely large upper critical fields in $Nb_2Pd_{0.81}S_5$

Q. Zhang[1], G. Li[1], D. Rhodes[1], A. Kiswandhi[1,2], T. Besara[1], B. Zeng[1], J. Sun[1], T. Siegrist[1,3], M. D. Johannes[4] & L. Balicas[1]

[1]National High Magnetic Field Laboratory, Florida State University, Tallahassee, Florida 32310, USA, [2]Department of Physics, Florida State University, Tallahassee, Florida 32306-3016, USA, [3]Department of Chemical and Biomedical Engineering, Florida State University, Tallahassee, Florida 32310, USA, [4]Center for Computational Materials Science, Naval Research Laboratory, Washington, DC 20375, USA.

Here, we report the discovery of superconductivity in a new transition metal-chalcogenide compound, i.e. $Nb_2Pd_{0.81}S_5$, with a transition temperature $T_c \cong 6.6$ K. Despite its relatively low $T_c$, it displays remarkably high and anisotropic superconducting upper critical fields, e.g. $\mu_0 H_{c2} (T \rightarrow 0 \text{ K}) > 37$ T for fields applied along the crystallographic $b$-axis. For a field applied perpendicularly to the $b$-axis, $\mu_0 H_{c2}$ shows a linear dependence in temperature which coupled to a temperature-dependent anisotropy of the upper critical fields, suggests that $Nb_2Pd_{0.81}S_5$ is a multi-band superconductor. This is consistent with band structure calculations which reveal nearly cylindrical and quasi-one-dimensional Fermi surface sheets having hole and electron character, respectively. The static spin susceptibility as calculated through the random phase approximation, reveals strong peaks suggesting proximity to a magnetic state and therefore the possibility of unconventional superconductivity.

The last couple of decades has seen the discovery of a series of new superconducting compounds displaying relatively low transition temperatures[1–3] ($T_c$) but whose properties are unique[1–3] and in some cases has led to the discovery of entire new families of materials[3]. In the case of $Sr_2RuO_4$ the intellectual motivation to study this low $T_c$ compound is due to experimental results suggesting the possibility of Cooper pairs pairing in a spin-triplet configuration[4,5] which is not subject to the so-called Clogston-Chandrasekhar paramagnetic limitation[6,7] under the application of an external magnetic field. As for the $Na_{0.3}CoO_2 \cdot yH_2O$ compound, most experimental results were interpreted as pointing towards a rather exotic gap whose wave-function presumably would be of $f$-wave symmetry leading also to a triplet state[8]. Finally, LaFePO[3] another low $T_c$ compound, led to the discovery of an entire new family of Fe-based superconductors with $T_c$s as high as 56 K[9] suggesting that additional families of high temperature superconductors with even higher $T_c$s remain to be discovered in new systems.

Here, we report the discovery of a new low dimensional transition metal-chalcogenide based compound, i.e. $Nb_2Pd_{0.81}S_5$ which shows remarkable physical properties and might become the basis of a new family of unconventional superconductors. In particular, this compound displays a remarkably large upper critical field relative to its transition temperature which surpasses by far the expected Pauli limiting field, and which is considerably larger than the one reported for the technologically relevant $Nb_3Sn$ compound ($\mu_0 H_{c2} \sim 30$ T, with $T_c = 18$ K)[10,11] which is widely used in superconducting applications such as the fabrication of superconducting magnets. Furthermore, its ratio of $\mu_0 H_{c2} (T \rightarrow 0 \text{ K})$ to $T_c$, is larger than those of the new Fe based superconductors, e.g. $\beta$-FeSe (20 T/8.7 K)[12], $Ba_{1-x}K_xFe_2As_2$ ($\sim$70 T/28 K)[13], and even higher than the ratio reported for the Chevrel-phase $PbMo_6S_8$ (60 T/13.3 K)[14] compound. The results discussed below suggest that this new compound is an unconventional superconductor, of perhaps triplet character.

## Results

$Nb_2Pd_{0.81}S_5$ crystallizes in the monoclinic space group $C2/m$, and the structure is shown in Fig. 1 a, b, and c. It is iso-structural to $Nb_2Pd_{0.71}Se_5$ which was previously described in Ref. 15. As seen in the figure, the structure is composed of sheets of Pd, Nb, and S along the b-axis, with trigonal prisms formed by S (in yellow) atoms coordinating the Nb sites (in blue) which are further linked by square planar Pd (in grey). As there are two sets of Pd and of Nb sites, each set forms a unique building block. The first building block consists of two $NbS_6$ trigonal prisms linked by square planar Pd and the second block is formed by two $NbS_7$ mono-capped trigonal prisms with the capping atoms shared by the neighboring prism, and prisms translated by $b/2$. Square planar Pd atoms link this second building block to the equivalent block in the adjacent layers. This Pd site shows partial occupation and here thereafter we will label this as the Pd(2) site. As discussed below, the level of partial occupation of the Pd(2) site has important consequences for the electronic properties of this compound. X-ray diffraction measurements shown in Fig. 1 d can be fit to the above crystallographic structure with no traces of impurity phases. Results of the Rietveld refinements are given in the supplementary data section.

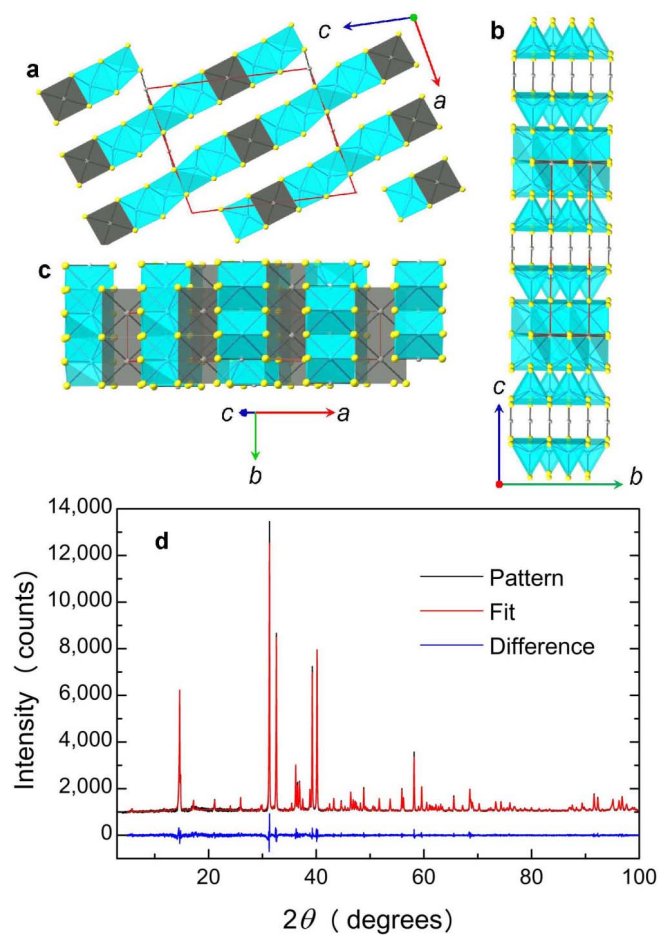

Figure 1 | **Crystallographic structure of $Nb_2Pd_{0.81}S_5$.** (a) Polyhedral representation of the crystallographic structure (orthorhombic space group $Cmcm$) of $Nb_2Pd_{0.81}S_5$ as seen from a perspective along the b-axis. S is depicted in yellow, Nb in turquoise and Pd in gray. S displays square pyramidal coordination around the Nb sites, while Pd shows a simple square-planar coordination. Notice, how these basic building blocks develop chain like structures. Red lines depict the boundaries of the first-Brillouin-zone. (b) Polyhedral representation of the crystallographic structure of $Nb_2Pd_{0.81}S_5$ as seen from a perspective along the a-axis. (c) same as in (a) or (b) but for a perspective nearly along the c-axis. (d) Powder x-ray diffraction as obtained from a bunch of randomly oriented single crystals, collected by using the Cu K$\alpha$ line. Black line corresponds to the actual x-ray diffraction data, red line is a fit to the structure previously reported in Ref. 15 and blue is the difference between both datasets, showing no evidence of impurities. The actual difference in height between measured and calculated peaks is due to slight discrepancies in the actual content of Pd.

Figures 2 a and b show respectively, the temperature dependence of the electrical resistivity and of the magnetic susceptibility, indicating a transition towards a zero resistivity state at $T_c \sim 6.5$ K, which is followed by a concomitant pronounced diamagnetic signal that is suppressed by the application of an external magnetic field. Figure 2 c shows the heat capacity C normalized by T as a function of temperature, under fields of 0 and 9 T, respectively, for a second batch of single crystals with slightly lower $T_c$ (according to the susceptibility data). For these measurements, hundreds of single crystals were pressed into a pellet although there are variations in stoichiometry between single-crystals from any given batch. Once the phononic contribution is subtracted, see Fig. 2 d, a clear anomaly is seen at the superconducting transition which is suppressed by the application of an external field of 9 T. These observations correspond to the standard response expected for bulk type II superconductors. The size of the anomaly, normalized with respect to the electronic contribution $\gamma$ to the heat-capacity above $T_c$ or $\Delta C/\gamma \approx (20.5-15)$ mJ/(molK$^2$)/15 mJ/(molK$^2$) = 0.37 is considerably smaller than the BCS value of 1.43, indicating that a sizeable fraction of the crystals might not be superconducting. Notice that small anomalies in the heat capacity have also been observed in single-crystals of Fe based superconductors[16].

$Nb_2Pd_{0.81}Se_5$ displays remarkably large superconducting upper critical fields, as seen in Figs. 3 a and b. For instance, compare $\mu_0 H^b_{c2}(T \rightarrow 0 K) \cong 37$ T for fields applied along the b-axis, with the expected weak coupling Pauli limiting field $H_p \sim 1.84 T_c \cong 12$ T, implying $\mu_0 H^b_{c2}(T \rightarrow 0 K)/H_p \sim 3$. For example, this ratio is comparable or surpasses the value reported for the strongly correlated heavy-fermion compound $CeCoIn_5$ for which $T_c = 2.3$ K[17] and $H^{ab}_{c2}(T \rightarrow 0 K) \sim 12$ T[18], implying a ratio of $\sim 2.8$. Such a large value for the upper critical fields necessarily implies a small coherence length $\xi = (2\pi\mu_0 H_{c2}(T \rightarrow 0 K)/\phi_0)^{1/2} \leq 30$Å. The high crystallographic and concomitant electronic anisotropy in $Nb_2Pd_{0.81}Se_5$ coupled to its extremely high upper critical fields, converts it into a prime candidate for the search of additional superconducting phases such as the Farrel-Fulde-Larkin-Ovchinnikov state[19,20] (claimed to have been observed in $CeCoIn_5$)[21], particularly if one could make this compound in nearly stoichiometric form.

Notice that the very large values of $H_{c2}(T \rightarrow 0 K)$ cannot be simply attributed to demagnetization factors associated to the geometry of our crystals. In effect, our crystals are needle-like, with a typical length l of a few millimeters, and cross sections A ranging from $10^{-12}$ m$^2$ to $10^{-10}$ m$^2$. The ratio l/A ranges from $\sim 100$ to 1000, thus these crystals can be approximated to very long cylinders. The demagnetization factor for very long (or infinite) cylinders is very small (or zero), meaning that along the axis of the cylinder (or along b-axis of our single crystals) the field inside the crystal, $H = H_{ext} + DM$, where D is the demagnetization factor and M is the magnetization of the sample, is basically the external field $H_{ext}$. The largest values of $H_{c2}(T \rightarrow 0 K)$ reported in this manuscript are precisely for fields applied along the b- or needle-axis. For fields applied perpendicularly to the axis of a long cylinder, D is $\sim 1/2$ meaning that H could be considerably smaller than $H_{ext}$. This implies that our measurements could overestimate $H_{c2}(T \rightarrow 0 K)$ for $H \perp b$, and therefore our samples would be effectively more anisotropic than our measurements indicate. The demagnetization factor is particularly relevant for the Meissner phase, or below the lower critical field where the magnetic-field lines are expelled from the interior of the sample. But for very large fields very close to $H_{c2}(T)$, the density of vortices is extremely large, i.e. H has penetrated throughout the entire sample leaving a negligible volume fraction of superconductivity. In fact, $H_{c2}(T)$ is understood as the field at which magnetic flux lines fill the entire volume of a metallic and no longer superconducting compound. Therefore, in an experimental situation like ours the role of D for H approaching $H_{c2}(T)$ should be irrelevant.

The resulting superconducting phase diagram for both field orientations is shown in Fig. 3. It was constructed by measuring the resistivity as a function of the temperature under modest magnetic fields, and by sweeping the field to high values at constant temperatures. Notice that for fields along the b-axis, $H^b_{c2}(T)$ does not seem to saturate as the temperature is lowered. Even more surprising, for fields applied perpendicularly to the needle axis of the crystal or the b-axis, $H^{\perp b}_{c2}(T)$ shows a linear dependence in temperature down to lowest temperatures, in sharp contrast to what is seen in most superconductors, i.e. a linear dependence for temperatures close to the superconducting critical temperature, followed by saturation, or a concave down curvature at lower temperatures. Such an anomalous temperature-dependence was recently observed in the new Fe based superconductors[13,21] and claimed to result from multi-band effects: superconducting gaps of different magnitudes open up on different Fermi surface sheets, each associated with bands of distinct electronic anisotropy. In fact, as seen in the lower panel of this figure, the superconducting anisotropy $\gamma = H^b_{c2}/H^{\perp b}_{c2} = \sqrt{m_b/m_{\perp b}}$ where $m$ is the band-dependent effective mass, is temperature dependent as is indeed the case for all Fe based pnictide/chalcogenide superconductors. This further suggests that $Nb_2Pd_{0.81}Se_5$ is a new multi-band superconductor.

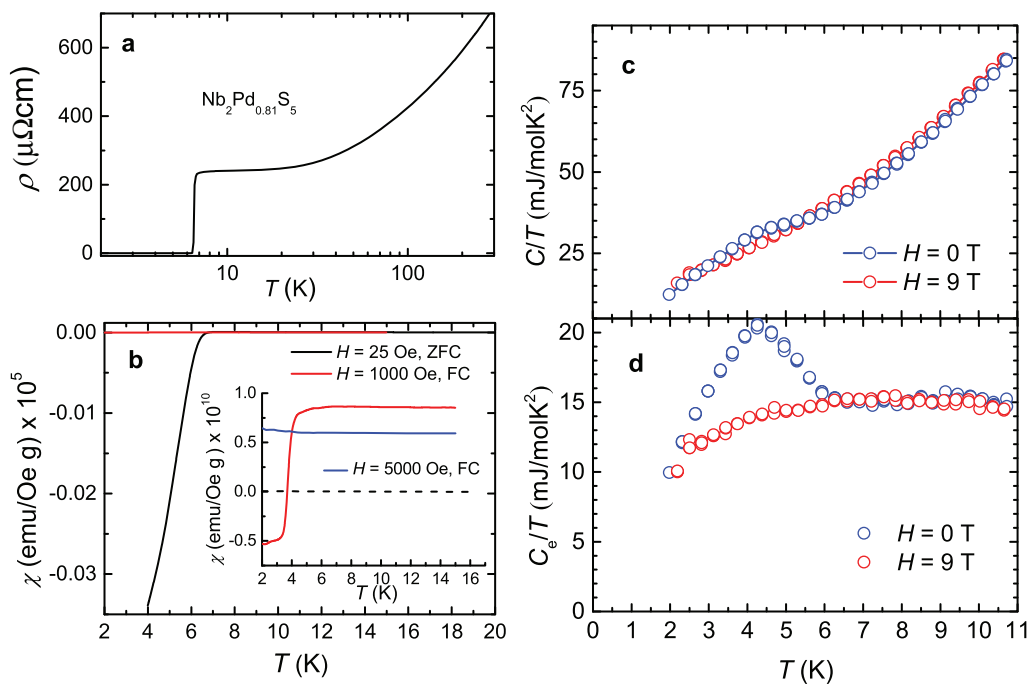

**Figure 2 | Electrical transport, magnetic susceptibility and heat capacity data.** (a) Resistivity as a function of temperature for $Nb_2Pd_{0.81}S_5$ single crystal, revealing a superconducting transition whose middle point of the resistive transition is located at $T_c \cong (6.5 \pm 0.2)$ K. (b) Magnetic susceptibility $\chi$ as function of temperature acquired either under a zero-field cooled condition for a field $H = 25$ Oe (black line), or under a field-cooled condition for $H = 1000$ Oe (red line). Inset: a detail of $\chi(T)$ at low temperatures obtained under a field of 1000 Oe, which still reveals evidence for superconductivity at $T \sim 4$ K. The diamagnetism is completely suppressed by the application of a field of 5000 Oe (blue line). No evidence for magnetic impurities, such as Curie-Weiss behaviour is observed in the magnetic susceptibility obtained under $H = 5000$ Oe. This susceptibility study was performed on a bunch containing hundreds of single-crystals; it confirms bulk superconductivity in this material. (c) Heat capacity as a function of temperature for a second synthesis batch containing hundreds of single crystals, under zero field and under 9 T, respectively. A large fraction of these crystals are of nominal $Nb_2Pd_{0.81}S_5$ composition, but this batch also contains a sizeable fraction of off-stoichiometry crystals as determined by EDS analysis. (d) Electronic contribution to the heat capacity, after subtraction of the phonon contribution, i.e. $\propto T^3$, showing an anomaly at the superconducting transition. Therefore, superconductivity in this system is bulk in nature.

## Discussion

So far, multi-band superconductivity has been reported in $MgB_2$[22] and in the Fe pnictices/chalcogenide superconductors[23,24]. The first is a phonon mediated superconductor, but the second are believed to possess a spin-fluctuations mediated pairing mechanism[25] due to the proximity of their superconducting phase to a magnetic state. In the Fe-based compounds as in the cuprates, superconductivity emerges by doping an antiferromagnetic parent compound which is metallic and a Mott insulator, respectively. This proximity to magnetism leads to an unconventional symmetry for the superconducting gap wave-function, i.e. $d$-wave symmetry for the cuprates[26], while the Fe-based compounds are believed, given the amount of experimental evidence, to possess the so-called $s\pm$ gap symmetry[27]. So, far our studies have not revealed any clear evidence for the proximity to a magnetic state in $Nb_2Pd_{0.81}Se_5$.

To shed some light about the possible superconducting pairing mechanism in $Nb_2Pd_{0.81}Se_5$ we have performed band structure calculations in order to determine both the geometry of the Fermi surface and the tendencies of this compound towards a magnetic instability, the results are summarized in Fig. 4. The Fermi surfaces and susceptibilities were calculated using the experimental lattice constants and atomic positions established in the present work. The centering symmetry of space group #12 ($C2/m$) was eliminated so that half the Pd at the 2a Wyckoff position could be removed, resulting in a doubled unit cell with formula $Nb_4Pd_{1.5}S_{10}$. Density functional theory calculations using Wien2K[27] with the Generalized Gradient Approximation (GGA)[28] to the exchange correlation potential were employed to calculate the self-consistent energy eigenvalues at 16,000 k-points in the reciprocal lattice. The doubled real space cell results in band-folding in the smaller reciprocal space cell. As seen in Fig. 4 a the resulting Fermi surface is composed of quasi-two-dimensional sheets (maroon surfaces) of hole character and strongly warped quasi-one-dimensional (Q1D) sheets with both electron (golden) and hole (blue) character. Therefore, given that the resulting Fermi surface is composed of multiple sheets, having both electron- and hole-character, this system can indeed be classified as a multi-band superconductor. This doubling of the unit cell has little effect on the Q1D surfaces, but it slightly distorts the 2D hole-like surfaces. The three dimensional like pockets seen in Fig. 4 a are electron-like with a three-dimensional appearance but are actually narrow slices through a slightly warped Q1D surface that is extremely sensitive to the Fermi energy, which is in turn sensitive to the exact Pd content of the sample. In order to mimic the role played by variations in Pd content, we shifted the Fermi energy by $-5$ meV which is found to eliminate these pockets entirely, while a shift of +12 meV converts them into highly one-dimensional sheets with strong nesting properties (compare Figs. 4 a and 4 c). These small numbers are nearly within the computational accuracy and within the certainty of the actual Pd content. We therefore use the positive shift to calculate the nesting function in order to illustrate the strong nesting contribution.

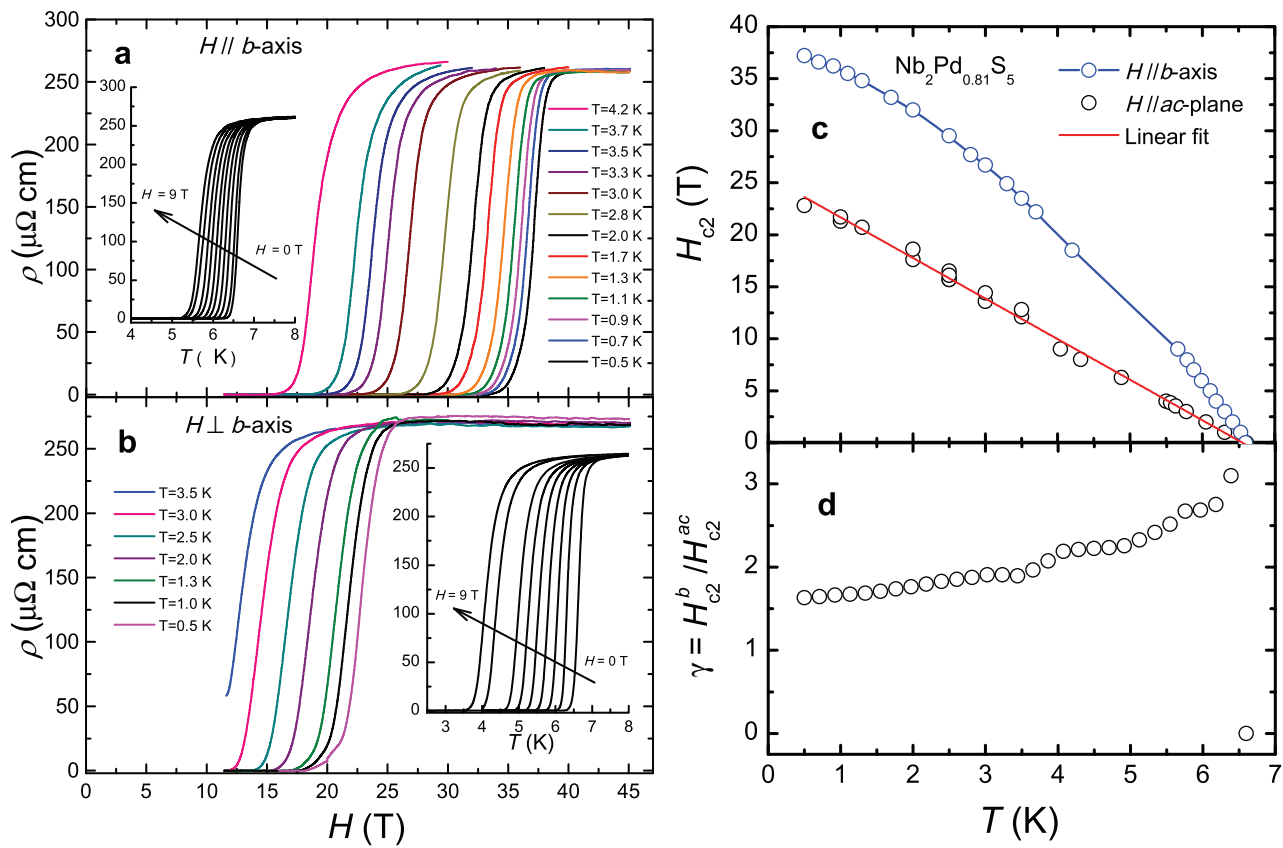

**Figure 3** | **Electrical transport, superconducting phase-diagram and anisotropy.** (a) and (b) Resistivity as a function of magnetic field for a $Nb_2Pd_{0.81}S_5$ single-crystal for several temperatures and respectively, for fields applied along the needle or the *b*-axis of the single crystal or perpendicularly to it. Both insets: Temperature dependence of the resistive transition for several fields applied along the b-axis (in (a)) and perpendicularly to it (b). (c) Superconducting upper critical field as a function of the temperature for $Nb_2Pd_{0.81}S_5$ and for both field orientations (as extracted from the data in (a) and (b)). Notice how anomalous is the behaviour of $H_{c2}(T)$ for fields applied perpendicularly to the b-axis, i.e. it shows a linear in temperature dependence. While for fields along the needle or the b-axis of the single crystal one does not observe a clear saturation of the upper critical fields at the lowest temperatures as observed in conventional superconductors. (d) anisotropy of the upper critical fields as a function of the temperature. Notice that the anisotropy is temperature dependent as observed in multi-band superconductors.

Figures 4 b and 4 e show the real part of the spin-susceptibilities $\chi(q)$ associated with the calculated Fermi surfaces shown in Figs. 4 a and 4 c, respectively. The ordering vector of a magnetic state is determined by peaks in the real part of $\chi(q)$ and this term also provides the pairing interaction for fluctuation mediated superconductivity. As seen, small shifts such as +12 meV, lead to pronounced spikes in $\chi(q)$ at small q values suggesting the proximity to an itinerant magnetically ordered state that would be characterized by long modulation wave-lengths. Using a fixed spin moment method to calculate the Stoner factor (I) reveals that only a very minor upward shift of the Fermi energy is required to cause a magnetic instability (NI = 0.98). Since the Pd content in our calculations is 0.75, while the measured content is 0.81, such a shift is expected to be a realistic approximation of the true electron count. By adding a small amount of extra negative charge to the calculation, very flat bands near $\varepsilon_F$ become the Q1D Fermi surface sheets seen in Figure 4 d. These give rise to strong nesting properties that translate into peaks in $\chi(q)$ while also raising the density of states sufficiently to stabilize ferromagnetic long range order (the peaks near to but not *at* zero in Figure 4 e suggest that the true ground state is likely to have a more complex long wavelength pattern). The existence of a magnetic state in calculations, while unseen in our experiments, is often an indication that the system is very near to magnetic long range ordering which is suppressed by spin fluctuations. The sensitive dependence on electron count also suggests that doping will be a fruitful avenue for manipulating the competition between magnetism and superconductivity.

These findings are rather intriguing, because the proximity to a magnetically ordered state in strongly correlated electronic systems, such as the cuprates, the heavy-fermion inter-metallic compounds, the organic superconducting compounds or the Fe based superconductors, has systematically led to proposals for unconventional superconducting pairing scenarios. Take for example, the quasi-one-dimensional organic conductors (e.g. $(TMTSF)_2X$, where TMTSF is an organic molecule that stands for tetramethyltetratiofulvalene and X is anion such as $PF_6^-$, $ClO_4^-$, etc.) which as in the present case, are also characterized by Q1D Fermi surface sheets and fairly large upper critical fields[29]. In this compound the combination of low dimensionality, relatively large upper critical fields, and sensitivity to non-magnetic impurities[30] was taken as possible evidence for triplet pairing[31,32]. For $Nb_2Pd_{0.81}Se_5$ we observed samples displaying lower superconducting transition temperatures, but at the moment it remains unclear if $T_c$ is affected by the amount of disorder in these crystals or if this decrease in $T_c$ is associated to stoichiometry induced changes in bandwidth and concomitant density of states at the Fermi level.

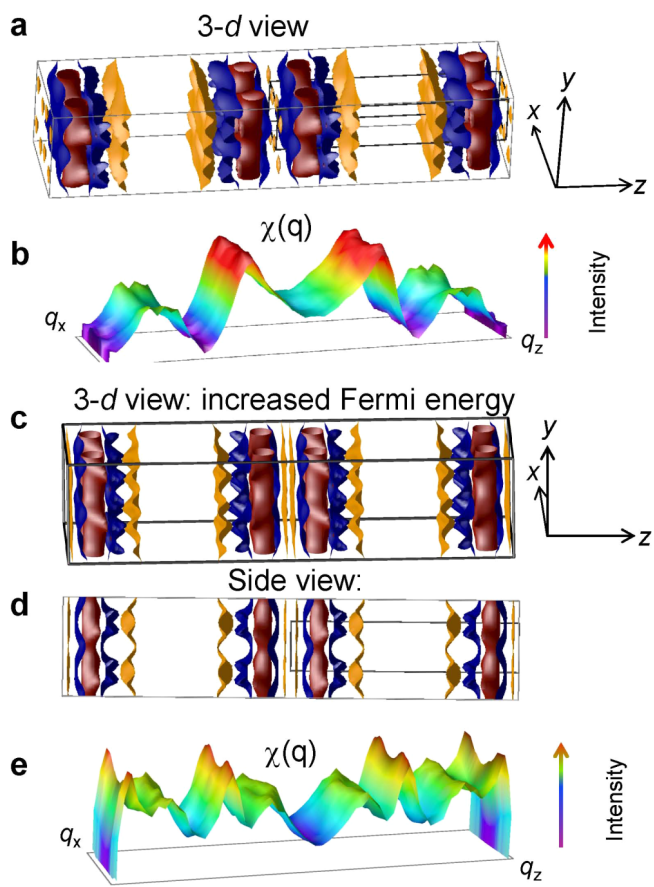

**Figure 4 | Calculated Fermi surface and spin susceptibility.** (a) The geometry of the Fermi surface in a $2 \times 2 \times 2$ conventional reciprocal unit cell for $Nb_2Pd_{0.75}S_5$ as obtained through DFT calculations. Maroon and blue FS sheets are hole-like, while gold pockets are of electron-like character. Notice that the Fermi surface is composed of both quasi-two-dimensional nearly cylindrical sheets as well as highly corrugated quasi-one dimensional ones. The small hole pockets appear to be three dimensional pockets but are actually just narrow slices through a slightly warped 1D surface that is extremely sensitive to the Fermi energy, which is in turn sensitive to the exact Pd content of the sample. (b) Real part of the spin response function $\chi(q)$ as a function of the planar reciprocal coordinates in a single conventional cell, calculated by using the experimental lattice constants and the atomic positions established in the present work for the fraction $x = 0.75$ of Pd. The strong peaks do not have their origin in Fermi surface nesting. (c) Increasing the Pd content to more realistic values such as $x = 0.81$ is equivalent to raising the Fermi energy which is found to stabilize rather flat quasi-one-dimensional electron-like Fermi surface sheets. (d) Side view of the Fermi surface sheets shown in (c) revealing a modest degree of warping. (e) $\chi(q)$ as a function of planar reciprocal coordinates. Note the emergence of pronounced nesting-driven spikes in $\chi(q)$ for small q-vectors, suggesting that this system is close to a long wavelength magnetic instability, or equivalently that pronounced spin-fluctuations are expected for this system.

Low-dimensionality, proximity to a magnetically ordered state, multi-band effects, and extremely high upper critical fields, are all ingredients pointing towards an unconventional superconducting state in $Nb_2Pd_{0.81}Se_5$. An important future consequence of this study is the possibility of synthesizing, through chemical substitution, i.e. Ta for Nb, Te for S, or Pt for Pd, new compounds displaying perhaps higher superconducting transition temperatures, or even higher upper critical fields due, for example, to an enhanced spin-orbit coupling.

## Methods

$Nb_2Pd_{0.81}S_5$ was grown through a solid state reaction by reacting under Ar atmosphere, Nb (99.99%), Pd (99.99%), and S (99.999%) powders in the ratio of 2 : 1 : 6. These mixtures were heated up to a peak temperature between 725 and 850°C at a rate of 10°C/h. The quartz ampoules containing the material were kept at this maximum temperature for 48 h and then quenched to room temperature. The obtained single crystals formed long needles, i.e. several millimetres in length, but having small cross sectional areas ranging from $2 \times 2$ μm² to approximately $10 \times 10$ μm². Single crystal diffraction on these individual fibres indicates strong twinning.

The stoichiometric composition was determined through energy dispersive x-ray spectroscopy. To determine the crystallographic structure a bunch of randomly oriented crystals were used as a means of extracting the powder x-ray diffraction pattern which revealed no impurity phases. Nevertheless, a few crystals selected for resistivity measurements revealed insulating, instead of metallic behaviour. At the moment it remains unclear if these crystals belong to a different phase or if small changes in stoichiometry might lead to an electronic/magnetic phase. The same bunch of crystals was used for DC susceptibility measurements performed through a commercial Squid magnetometer. A conventional four terminal configuration was used for resistivity measurements which were performed at low fields either by using a commercial Physical Parameter Measurements System or the 45 T hybrid magnet coupled to a ³He cryogenic system.

1. Maeno, Y. *et al.* Superconductivity in a layered perovskite without copper. *Nature* **372**, 532 (1994).
2. Takada, K., Sakurai, H., Takayama-Muromachi, E., Izumi, F., Dilanian, R. A. & Sasaki, T. Superconductivity in twodimensional $CoO_2$ layers. *Nature* **422**, 53 (2003).
3. Kamihara, Y. *et al.* Iron-based layered superconductor: LaOFeP. *J. Amer. Chem. Soc.* **128**, 10012 (2006).
4. Ishida, K. *et al.* Spin-triplet superconductivity in $Sr_2RuO_4$ identified by O-17 Knight shift. *Nature* **396**, 658–660 (1998).
5. Luke, G. M. *et al.* Time-reversal symmetry breaking superconductivity in $Sr_2RuO_4$ *Nature* **394**, 558–561 (1998).
6. Clogston, A. M. Upper Limit for the Critical Field in Hard Superconductors. *Phys. Rev. Lett.* **9**, 266 (1962).
7. Chandrasekhar, B. S. A Note on Maximum Critical Field of High-field Superconductors. *Appl. Phys. Lett.* **1**, 7 (1962).
8. Mazin, I. I. & Johannes, M. D. A critical assessment of the superconducting pairing symmetry in $Na_xCoO_2.yH_2O$. *Nat. Phys.* **1**, 91–93 (2005).
9. See, for example, Liu, R. H. *et al.* Anomalous transport properties and phase diagram of the FeAs-based $SmFeAsO_{1-x}F_x$ superconductors. *Phys. Rev. Lett.* **101**, 087001 (2008).
10. Matthias, B. T., Geballe, T. H., Geller, S. & Corenzwit, E. Superconductivity of $Nb_3Sn$. *Phys. Rev.* **95**, 1435–1435 (1954).
11. Scanlan, R. M., Malozemoff, A. P. & Larbalestier, D. C. P. Superconducting materials for large scale applications. *P. IEEE* **92**, 1639–1654 (2004).
12. Lei, H. *et al.* Multiband effects on β-FeSe single crystals. *Phys. Rev. B* **85**, 094515 (2012).
13. Yuan, H. Q. *et al.* Nearly isotropic superconductivity in $(Ba,K)Fe_2As_2$. *Nature* **457**, 565–568 (2009).
14. Okuda, K., Kitagawa, M., Sakakibara, T. & Date, M. Upper Critical Field Measurements up to 600 kG in $PbMo_6S_8$. *J. Phys. Soc. Jpn.* **48**, 2157–2158 (1980).
15. Keszler, D. A., Ibers, J. A., Shang, M. & Lu, J., Ternary and Quaternary Transition-Metal Selenides: Syntheses and Characterization. *J. Solid State Chem.* **57**, 68 (1985).
16. Welp, U. *et al.* Calorimetric determination of the upper critical fields and anisotropy of $NdFeAsO_{1-x}F_x$ single crystals. *Phys. Rev. B* **78**, 140510 (2008).
17. Petrovic, C. *et al.* Heavy-fermion superconductivity in $CeCoIn_5$ at 2.3 K. *J. Phys.-Condens. Mat.* **13**, L337–L342 (2001).
18. Bianchi, A., Movshovich, R., Capan, C., Pagliuso, P. G. & Sarrao, J. L. Possible Fulde-Ferrell-Larkin-Ovchinnikov Superconducting State in $CeCoIn_5$. *Phys. Rev. Lett.* **91**, 187004 (2003).
19. Fulde, P. & Ferrell, R. A. Superconductivity in strong spin-exchange Field. *Phys. Rev.* **135**, A550 (1964).
20. Larkin, A. I. & Ovchinnikov, Y. N. Inhomogeneous State of Superconductors. *Zh. Eksp. Teor. Fiz.* **47**, 1136 (1964) [*Sov. Phys. JETP* **20**, 762 (1965)].
21. Yamamoto, A. *et al.* D. Small anisotropy, weak thermal fluctuations, and high field superconductivity in Co-doped iron pnictide $Ba(Fe_{1-x}Co_x)_2As_2$. *Appl. Phys. Lett.* **94**, 062511 (2009).
22. See, for example, Szabo, P. *et al.* Evidence for two superconducting energy gaps in $MgB_2$ by point-contact spectroscopy. *Phys. Rev. Lett.* **87**, 137005 (2001).
23. Hashimoto, K. *et al.* Microwave Penetration Depth and Quasiparticle Conductivity of $PrFeAsO_{1-y}$ Single Crystals: Evidence for a Full-Gap Superconductor. *Phys. Rev. Lett.* **102**, 017002 (2009).
24. Ding, H. *et al.* Observation of Fermi-surface-dependent nodeless superconducting gaps in $Ba_{0.6}K_{0.4}Fe_2As_2$. *EPL* **83**, 47001 (2008).
25. See, for example, Lee, P. A., Nagaosa, N. & Wen, X. G. Doping a Mott insulator: Physics of high-temperature superconductivty. *Rev. Mod. Phys.* **78**, 17 (2006), and references therein.

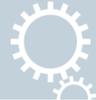

26. Mazin, I. I., Singh, D. J., Johannes, M. D. & Du, M. H. Unconventional Superconductivity with a Sign Reversal in the Order Parameter of LaFeAsO$_{1-x}$F$_x$. *Phys. Rev. Lett.* **101**, 057003 (2008).
27. Blaha, P., Schwarz, K., Madsen, G. K. H., Kvasnicka, D. & Luitz, J. *WIEN2k, An Augmented Plane Wave + Local Orbitals Program for Calculating Crystal Properties* (Karlheinz Schwarz, Techn. Universitat Wien, Austria), 2001. ISBN 3-9501031-1-2.
28. Perdew, J. P., Burke, S. & Ernzerhof, M. Generalized Gradient Approximation Made Simple. *Phys. Rev. Lett.* **77**, 3865 (1996).
29. Lee, I. J., Naughton, M. J., Danner, G. M. & Chaikin, P. M. Anisotropy of the upper critical field in (TMTSF)$_2$PF$_6$. *Phys. Rev. Lett.* **78**, 3555 (1997).
30. Choi, M. Y., Chaikin, P. M. & Greene, R. L. *Phys. Rev. B* **34**, 7727 (1984); Choi, M. Y. *et al.*, ibid. **25**, 6208 (1982).
31. Abrikosov, A. A. Superconductivity in a quasi-one-dimensional metal with impurities. *J. Low Temp. Phys.* **53**, 359 (1983).
32. Gor'kov, L. P. & Jerome, D. Back to the problem of the upper critical fields in organic superconductors. *J. Phys. (Paris) Lett.* **46**, L-643–L681 (1985).

**Acknowledgements**

LB acknowledges financial support from DOE, Basic Energy Sciences, contract Nº DE-SC0002613. The NHMFL is supported by NSF through NSF-DMR-0654118, the State of Florida and DOE.

**Author contributions**

Q.Z. synthesized the single crystals of Nb$_2$Pd$_{0.81}$S$_5$. G.L., Q.Z., D.R. and L.B. performed transport measurements at low and high fields and analyzed the corresponding data. Q.Z. and B. Z. performed heat capacity measurements. A.K., T.B. and T.S. performed the x-ray structural analysis. Q.Z. and L.B. performed the magnetization measurements. M.D.J. performed the band structure, Fermi surface geometry and spin susceptibility calculations. L.B. conceived the project and wrote the manuscript including the input of all co-authors.

**Additional information**

**Competing financial interests:** The authors declare no competing financial interests.

**How to cite this article:** Zhang, Q. *et al.* Superconductivity with extremely large upper critical fields in Nb$_2$Pd$_{0.81}$S$_5$. *Sci. Rep.* **3**, 1446; DOI:10.1038/srep01446 (2013).